\documentclass[aps,prx,twocolumn,preprintnumbers,amsmath,amssymb,nofootinbib]{revtex4-2}
\usepackage{epsfig}
\usepackage{epstopdf}
\usepackage{lipsum}
\usepackage{multirow}
\usepackage{boldline}
\usepackage{xcolor}

\begin{document}

\title{Towards fractal origins of the community structure in complex networks:\\a model-based approach}
\author{Mateusz Samsel, Kordian Makulski, Michał \L epek, Agata Fronczak, Piotr Fronczak}
\affiliation{Faculty of Physics, Warsaw University of Technology, Koszykowa 75, PL-00-662 Warsaw, Poland}
\date{\today}

\begin{abstract}
In this paper, we pose a hypothesis that the structure of communities in complex networks may result from their latent fractal properties. This hypothesis is based not only on the general observation that many real networks have multilevel organization, which is reminiscent of the geometric self-similarity of classical fractals. Quantitative arguments supporting this hypothesis are: first, many non-fractal real complex networks that have a well-defined community structure reveal fractal properties when suitably diluted; second, the scale-free community size distributions observed in many real networks directly relate to scale-invariant box mass distributions, which have recently been described as a fundamental feature of fractal complex networks. We test this hypothesis in a general model of evolving network with community structure that exhibits dual scale invariance: at the level of node degrees and community sizes, respectively. We show that, at least in this model, the proposed hypothesis cannot be rejected. The argument for this is that a kind of fractal core can be identified in the networks studied, which appears as a macroscopic connected component when the edges between modules identified by the community detection algorithm are removed in a supervised manner.
\end{abstract}

\maketitle

\section{Introduction}
\label{sec:introduction}

Much has changed in our perception of nature over the past half century, since Mandelbrot first coined the term \textit{fractal} \cite{1982bookMandelbrot} to describe geometric objects that can be subdivided into parts, each of which is (at least approximately) a reduced-size copy of a whole. Nowadays, equipped with basic understanding of the fractal geometry \cite{1988bookFeder, 2009BundeEncyclopedia} one can see fractals almost everywhere. For this reason, when over twenty years ago the pioneers of the network science argued that real-world networks \textit{are neither regular nor completely random} \cite{1999BarabasiScience, 2003bookDorogovtsev, 2007bookCaldarelli}, it was natural to assume that most of them must also exhibit fractal properties. 

The belief in the fractal nature of complex networks was based on simple reasoning: Most real-world networks are characterized by different scale-invariant distributions (e.g. by the power-law-like node degree distribution), which is referred to as the \textit{scale-free property} of complex networks \cite{2018bookNewman, 2022bookDorogovtsev}.Thus, it has been speculated that since the geometric self-similarity of classical fractals is a special case of the broader mathematical concept of scale-invariance \cite{1997bookDubrulle, 2006bookSornette}, the scale-freeness of complex networks may appear due to their inherent fractality. These speculations seemed all the more plausible in light of the discovery of a hierarchical community structure bearing the hallmarks of geometric self-similarity in many real-world networks \cite{2002OltvaiScience, 2002RavaszScience, 2003RavaszPRE, 2004NewmanPRE, 2005PallaNature, 2006NewmanPNAS, 2007SalesPNAS}. Soon after, it was actually shown that some real networks (e.g. WWW and protein networks) have fractal  \cite{2005SongNature, 2006SongNatPhys, 2009RozenfeldEncyclopedia} or even multifractal \cite{Wang_2012,Liu_2019} properties. Today, however, from the perspective of more than twenty years of research on complex networks, in light of the successes of \textit{the fractal geometry of nature}, it is quite surprising that  fractal complex networks represent a really small fraction of all networks that have been studied so far \cite{2020bookRosenberg, 2021BogunaNatPhys}. In this paper we deliberate on this state of affairs. More precisely, we put forward a working hypothesis that the community structure, which (unlike the rarely observed fractality) characterizes many real-world networks, can be treated as a superstructured fractality that has been overwhelmed by the addition of connections leading to improved network functionality.  We show that this hypothesis cannot be rejected in a simple model of an evolving modular network, providing an interesting starting point for further research on real networks with community structure.

The hypothesis we intend to test is based not only on the general observation that many real-world networks have a multilevel organization, which brings to mind the geometric self-similarity of a classical fractal. There are certain observations of a quantitative nature behind it. First, many non-fractal real complex networks that do have a well-defined community structure, reveal fractal properties when properly thinned out, e.g. by removing less significant edges of small weight. Examples of such networks come from various fields and represent social, biological, and technological systems (see e.g. DBLP \cite{2022FronczakSciRep, 2023FronczakArxiv} and IMBD \cite{2013GallosPLoS} collaboration networks, functional network of the human brain \cite{2012GallosPNAS}, internet \cite{2006GohPRL, 2007KimPRE}). Second, the scale-free community size distributions observed in a number of real networks (see e.g. \cite{2003GuimeraPRE, 2004ClausetPRE, 2005PallaNature, 2007DanonInbook, 2009LancichinettiNewJPhys,2015EPJDataSci}) directly relate to the scale-invariant box mass distributions that have recently been described as the fundamental feature of fractal complex networks \cite{2023FronczakArxiv}.


In what follows, to test the hypothesis on the fractal origins of the community structure, in Sec.~\ref{SecModel}, we introduce and study a generic model of an evolving network that accounts for the scale-free heterogeneity of both node degrees and community sizes. In Sec.~\ref{SecFractal}, we show that at low densities of inter-module connections, networks generated using this model exhibit fractal properties, which, disappear as the density of these connections increases leaving the community structure as a remnant. Then, we discuss the method of filtering out the network connections in order to recover the underlying fractal core of the studied networks. The paper ends in Sec.~\ref{SecSummary} with a summary of the obtained results and a discussion of their consequences.

\section{Benchmark network model for hypothesis testing}\label{SecModel}
 
\begin{figure}[t]
	\includegraphics[width=0.6\columnwidth]{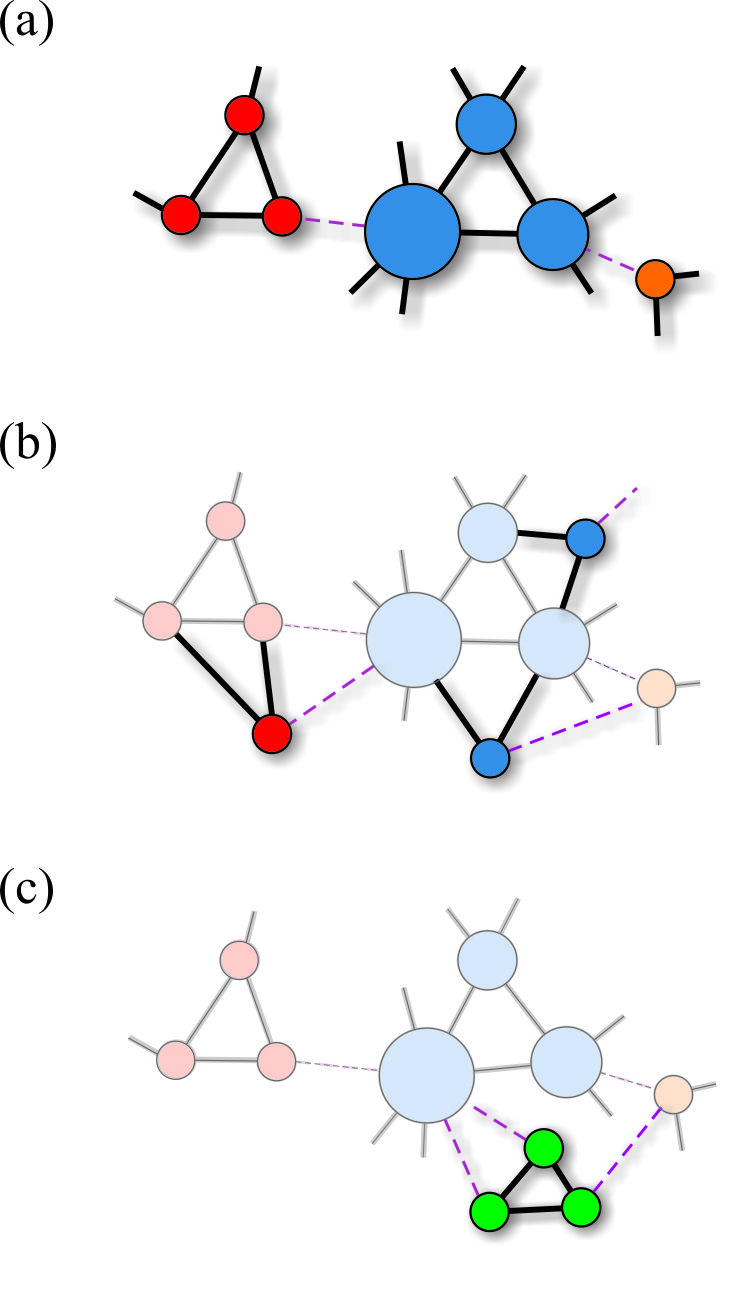}
	\caption{Illustration of the construction procedure of the considered benchmark network model with $a=2$ and $b=1$. The following parts of the figure show: (a) a fragment of the network in a certain time step $t$, and two modes, (b) and (c), of the network growth that could happen in the next time step $t+1$, which correspond to the expansion of existing groups and the creation of a new group, respectively. The nodes belonging to the same group are marked with the same color. The $A$-edges (inside the groups) are marked with solid lines, and the $B$-edges (between the modules) with dashed lines.\label{fig1}}
\end{figure}

\begin{figure}[t]
	\includegraphics[width=1.0\columnwidth]{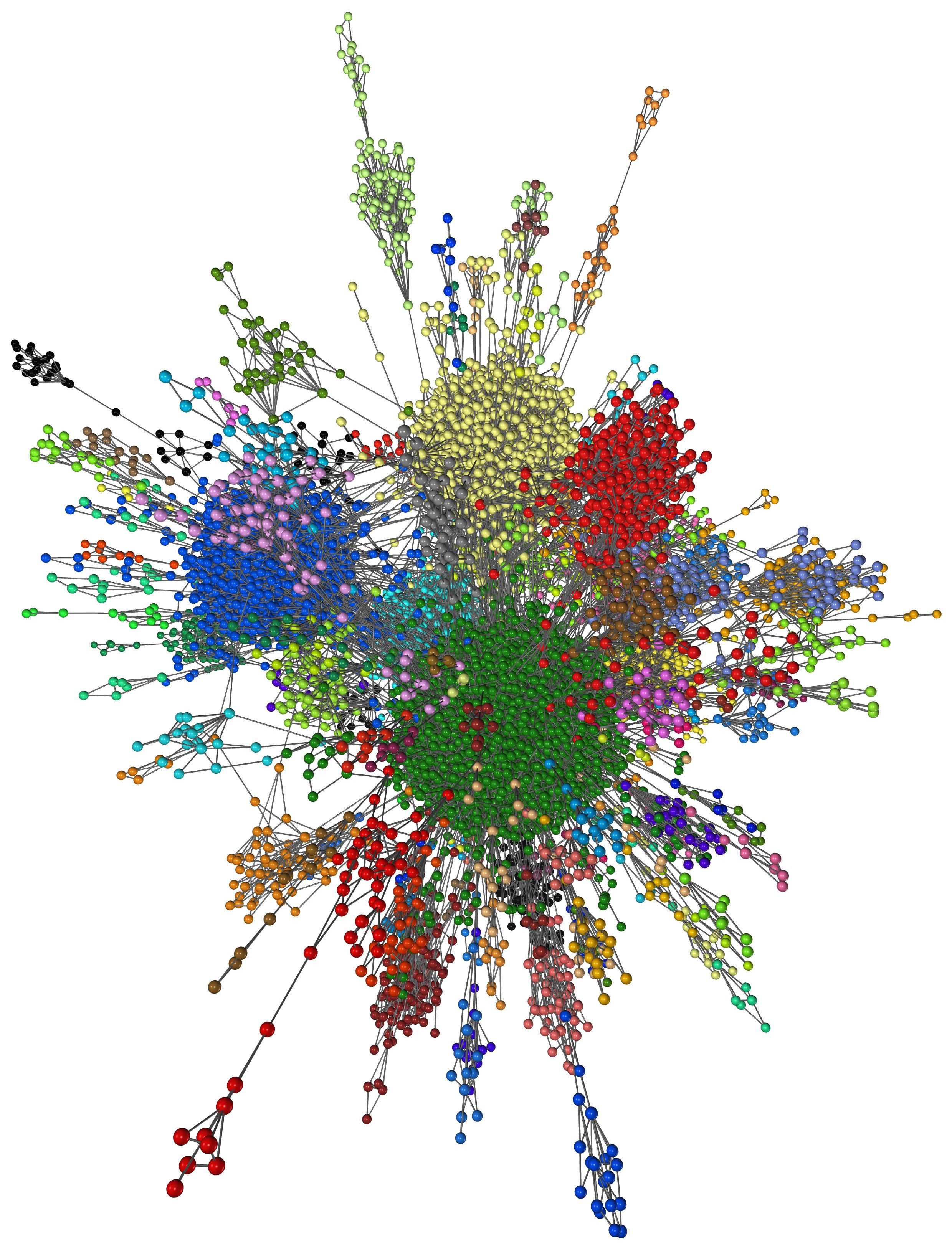}
	\caption{A realization of the benchmark network model with $N=10^4$ nodes for $p=0.8$, $a=2$ and $b=0.05$. Nodes that were assigned to the same group during network construction are marked with the same color.\label{fig2}}
\end{figure}

\begin{figure}[t]
	\includegraphics[width=1.0\columnwidth]{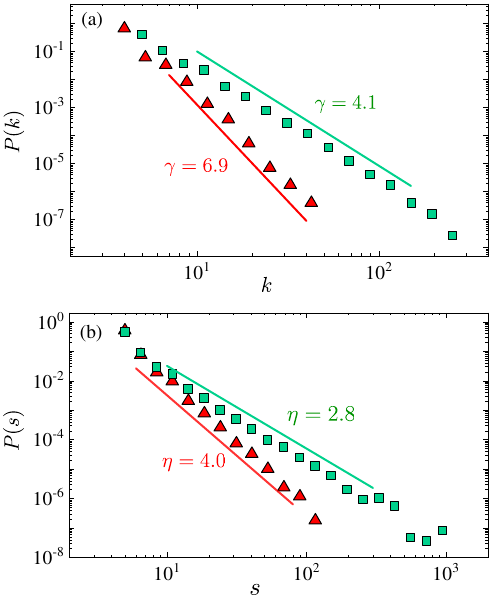}
	\caption{(a) Node degree distributions $P(k)$, and (b) group size distributions, $P(s)$, obtained as a result of numerical simulations of the benchmark networks with the following model parameters: $p=0.35$, $a=4$, $b=1$ (green squares), and $p=0.2$, $a=4$, $b=0.01$ (red triangles). The straight lines represent theoretically predicated slopes of the distributions given by Eqs.~(\ref{gamma0}) and (\ref{eta0}), respectively.\label{fig3}}
\end{figure}

\subsection{Motivation} 

The network model examined in this paper is inspired by models based on the preferential attachment rule (PAR), according to which the probability for the newly created node to establish connections to existing nodes depends on their degrees (e.g. the famous BA networks \cite{1999BarabasiScience, 1999BarabasiPhysA} and other related network models \cite{2004BarratPRL, 2007XiePhysA, 2009XuPhysA, 2013NicosiaPRL}). What distinguishes the model under consideration, especially in comparison with various benchmark charts for testing community detection algorithms \cite{2008LancichinettiPRE, 2009LancichinettiPRE, 2013FronczakPRE, 2017KowalczykPhysA, 2016YangSciRep}, is its evolving construction procedure, which uses dual PAR at the node and community levels, respectively. Both mechanisms have been observed in the evolution of real-world networks \cite{2003JeongEPL, 2006PollnerEPL, 2008LescovecPAR}. Although compared to the aforementioned benchmark charts, which are static and have predetermined properties, our model has the disadvantage that the characteristic exponents of the resulting distributions are not easy to control, its realistic construction procedure compensates for these shortcomings. This feature of our benchmark model is particularly important in the context of fractal networks, especially since the known models of such networks are either deterministic or boil down to the recursive reconstruction of the feature of geometric self-similarity \cite{2006SongNatPhys, 2007RozenfeldNewJPhys, 2015KuangSciChina, 2022YakuboPLOS, 2022MolontayInProc},  

\subsection{Construction procedure}\label{SECprocedure}

In our model, communities are defined as non-overlapping groups of nodes, meaning that each node $i$ can belong to only one group, and this membership is determined at the time of its birth, $t_i$. Since nodes belong to specific groups, their edges can be of two types: intra- and inter-module, which we refer to as $A$ and $B$ edges, respectively. Correspondingly, the degree of node $i$ is the sum of its internal and external degrees: $k_i=k_i^A+k_i^B$. 

The construction procedure of the model is as follows: The network starts to grow from an $a$-regular graph of $g$ nodes. (Next we will assume that $g=a+1$, which reduces the number of model parameters and simplifies analytical calculations, see Appendix). The seed nodes constitute the initial group. Then, at each subsequent time step, $g$ new nodes are added, which, with probability $p$ join existing groups, or form a new group. In the case when the existing groups are expanded (with the probability $p$), each of the newly added nodes can join a different group. The target groups are chosen preferentially (the larger the total degree of the group, the greater the probability of attracting a new node), and each newcomer creates $a$ preferential connections within its own group and $b$ preferential connections within the entire network. Otherwise, when the new nodes form a new group (with the probability $1-p$), the group is created as a clique of size $g$, with each node additionally creating $b$ preferential connections within the entire network. 

The two complementary modes of the network growth are schematically illustrated in Fig.~\ref{fig1}, where we assumed that the model parameters $a$ and $b$ are natural numbers, although this is not a necessary condition. In particular, this remark applies to the parameter $b$, which is later assumed to represent the \textit{average} number of $B$-edges that each newly added node creates. Accordingly, in Fig.~\ref{fig2} an example of the network obtained using this procedure with $b<1$ is shown. 

\subsection{Double scale-freeness}\label{doubleP}

The above construction procedure leads to networks with community structure characterized by the average mixing parameter $\mu=\left\langle\sum_i k_i^B/\sum_i k_i\right\rangle=b/(a+b)$ \cite{2008LancichinettiPRE, 2009LancichinettiPRE} and scale-free distributions of node degrees, 
\begin{equation}\label{Pk0}
P(k)\propto k^{-\gamma},
\end{equation}
and group sizes, 
\begin{equation}\label{Ps0}
P(s)\propto s^{-\eta}, 
\end{equation}
which are the prerequisites for testing our hypothesis. The characteristic exponents of these distributions (see Fig.~\ref{fig3}) are given by the model parameters:
\begin{equation}\label{gamma0}
	\gamma=2+\frac{a+b}{pa+b}\;\stackrel{b\ll a}{\simeq}\;2+\frac{1}{p}, 
\end{equation} 
and
\begin{equation}\label{eta0}
	\eta=1+\frac{pa+a+2b}{p(2a+b)+b}\;\stackrel{b\ll a}{\simeq}\;\frac{3}{2}+\frac{1}{2p}.
\end{equation} 
The theoretical derivations underlying Eqs.~(\ref{Pk0})–(\ref{eta0}) are similar to the continuous-time mean-field method used to determine the node degree distribution in the famous BA model \cite{1999BarabasiPhysA}. For this reason, since this method is widely known, in the Appendix, we only provide basic steps of the method, indicating and appropriately commenting only on those equations that differ significantly in both models.

\begin{figure}[t]
	\includegraphics[width=1.0\columnwidth]{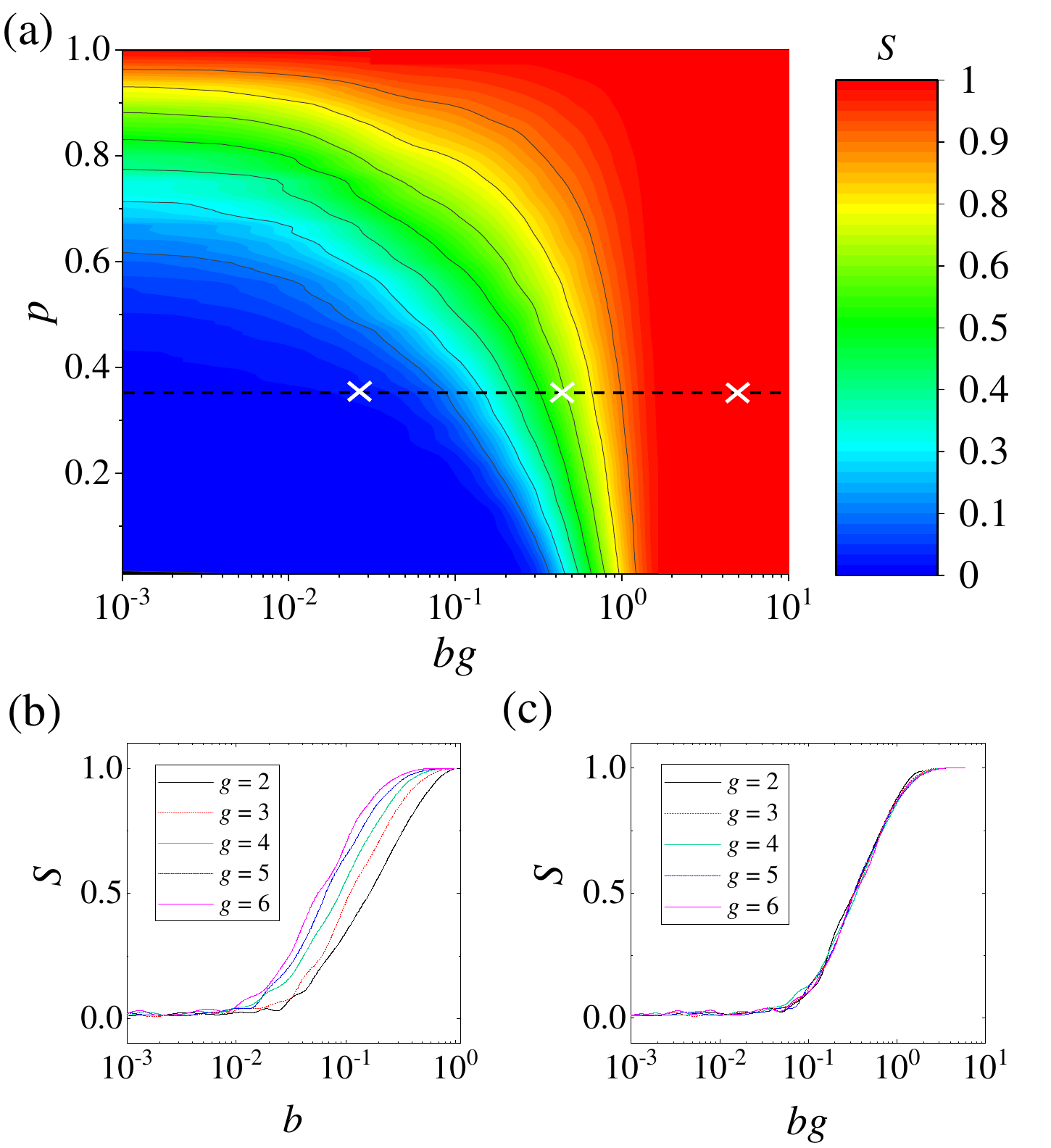}
	\caption{(a) Phase diagram for percolation transition in the considered benchmark model for $a\geq 1$. The areas marked with the same color represent the range of model parameters $(gb,p)$ for which the relative size of the giant component $S$ is fixed. (b,\;c) Relative size of the giant component $S$ as a function of the parameter $b$ for $p=0.35$ and different values of the parameter $g=a+1$. \label{fig4}}
\end{figure}

\subsection{Percolation}\label{percolation}

An important characteristic of the considered network model, affecting our further analyses, is that for $p\neq 1$ and sufficiently low densities of inter-group connections the networks may consist of many separate clusters. (Note that the case $p=1$ corresponds to BA networks.) In fact, there is a percolation transition in the model, with the threshold depending only on the probability $p$ and on the product $gb$ (meaning the average number of $B$-connections that each new group creates at birth), with $A$-connections of any density. The observed lack of direct dependence of the phase diagram on the density of intra-module connections means that at the percolation threshold, the studied networks are not trees, as is the case with networks without a community structure \cite{2018bookNewman, 2022bookDorogovtsev}, i.e. this means that the transition occurs at the mesoscopic (inter-group) level.

Phase diagram of the transition, which is shown in Fig.~\ref{fig4}, is  mainly for demonstration purposes. In the following, when studying the relation between community structure and fractality, we \textit{only} focus on the giant components (GCs) of these networks. Accordingly, the distributions $P(k)$ and $P(s)$, as well as the other quantities analyzed in the rest of the article, refer to these components, not to the network as a whole.

\subsection{Community structure}\label{Louvain}

Throughout this paper, community structure of the considered networks is identified using the Leiden algorithm \cite{2019TraagSciRep}, which is an improved version of the famous Louvain algorithm \cite{2008BlondelJStatMech}.  The algorithm is interesting in its self, especially in the context of fractality of complex networks, because it strongly relates to the procedure for renormalizing fractal networks, which we will discuss in the next section (see Sec. ~\ref{fractality}) and which is traditionally used to reveal the geometric self-similarity of networks. 

In short, the algorithm is a multistep technique based on a local optimization of the \textit{modularity}, which is the quantity that measures the density of links inside communities as compared to links between communities \cite{2004NewmanPRE}. In the first step, the algorithm finds communities by optimizing modularity locally on all nodes. The communities defined in this way are then grouped into a single node, and the first step is repeated. 

As expected due to the known high efficiency of this algorithm, the structure of the community discovered with its help corresponds very well with the original structure of the groups that arises as a result of the construction procedure of the considered model. Indeed, over a wide range of model parameters, up to the mixing parameter $\mu\simeq 0.8$, the normalized mutual information of both partitions (groups vs. communities) \cite{2005DanonJStatMech} does not fall below the value of 0.8, resulting in overlapping scale-free distributions of group and community sizes, i.e. $P(s)$ and $P(c)$, respectively. 

Later in the paper, the internal and external connections of the communities identified by the Leiden algorithm are called $\mathcal{A}$ and $\mathcal{B}$ edges, respectively, thus indicating their correspondence to the division into $A$ and $B$ edges as introduced by the construction procedure of the model (see Sec.~\ref{SECprocedure}).

\section{Fractal core underlying the community structure}\label{SecFractal}

\subsection{Factality in complex networks}\label{fractality}

In complex networks, fractality is traditionally assessed using the procedure of covering the network with non-overlapping boxes, with the maximum distance between any two nodes in each box less than the diameter $l_B$. More precisely, as defined by Song et al.~\cite{2005SongNature}, fractal complex networks exhibit power-law scaling:
\begin{equation}\label{defdB}
	N_B(l_B)\propto l_B^{-d_B},
\end{equation}
where $N_B(l_B)$ is the number of boxes of a given diameter, and $d_B$ is the fractal (or box) dimension. This power-law scaling implies that the average mass of a box (which is the average number of nodes belonging to such a box), also scales according to the power-law: 
\begin{equation}
	\langle m(l_B)\rangle=\frac{N}{N_B(l_B)}\propto l_B^{\;d_B}.
\end{equation}
The above scaling relation, however, says nothing about the distribution of masses of all the boxes used in the box-covering method which leads to Eq.~(\ref{defdB}). This problem has recently been addressed in Ref.~\cite{2023FronczakArxiv}, where it was shown that, in fractal complex networks, just like node degree distributions $P(k)$, the mass box distributions are also scale-free, 
\begin{equation}\label{Pm}
	P(m)\propto m^{-\delta},
\end{equation}
with the characteristic exponent $\delta$ independent on the diameter $l_B$ of the boxes with which one covers the network. Furthermore, in Ref.~\cite{2023FronczakArxiv} it was shown that the scaling relations (\ref{defdB}) and (\ref{Pm}) arise from the geometric self-similarity of fractal networks which manifests itself not only when comparing the microscopic structure of different boxes with a fixed diameter, but also when comparing the original network and its renormalized counterpart, which emerges when nodes belonging to the same box in the original network are replaced by a supernode in its renormalized version \cite{2005SongNature}. Indeed, the renormalization procedure just mentioned (which, according to the authors of the Louvain algorithm, inspired their method, Sec.~\ref{Louvain}) leaves not only the node degree distribution unchanged, but also the box mass distribution.

\subsection{Fractality in the benchmark model}

\begin{figure}[t]
	\includegraphics[width=0.88\columnwidth]{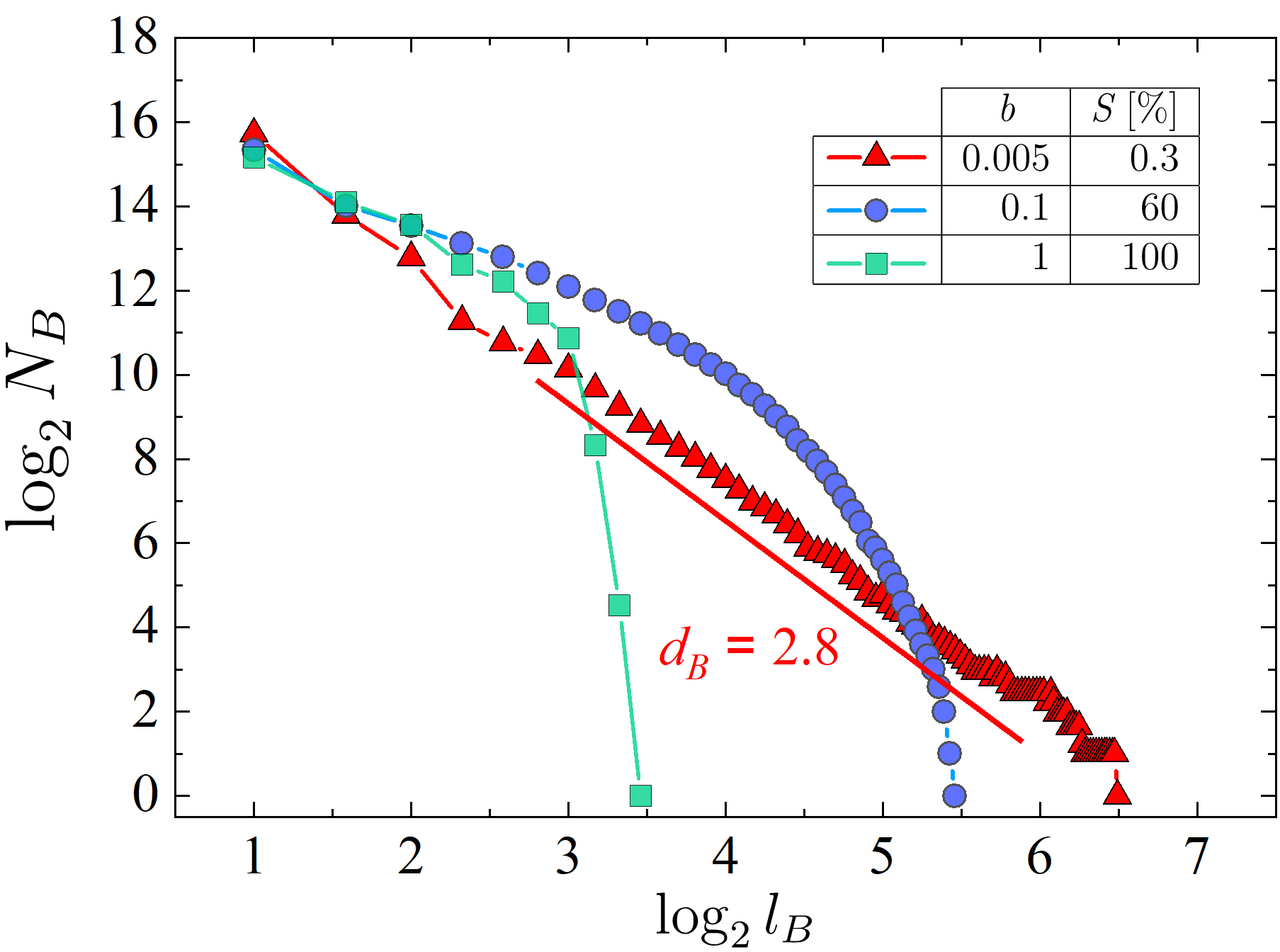}
	\caption{Box counting analysis of the considered network model for $p=0.35$, $a=4$ and different values of the parameter $b$, which correspond to different relative sizes $S$ of the giant components. The data series shown correspond to model parameters marked with white crosses in the phase diagram, Fig.~\ref{fig4}~(a).  \label{fig6}}
\end{figure}

\begin{figure}[t]
	\includegraphics[width=0.90\columnwidth]{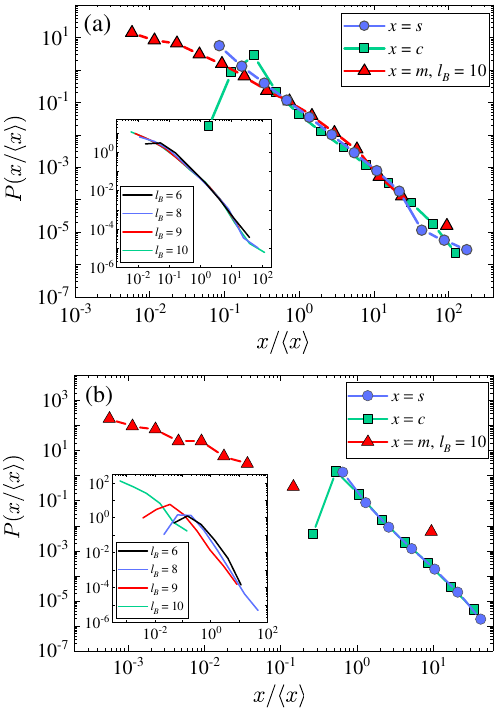}
	\caption{Comparison of various distributions $P(x)$, where $x\in\{s,\,c,\,m\}$, characterizing the mesoscopic structure of the studied networks for $p=0.35$, $a=4$ and two different values of $b$, corresponding to: (a) fractal network at the percolation threshold (with $b=0.005$) and (b) non-fractal network far from the threshold (with $b=1$) (cf. Fig.~\ref{fig6}). The insets in both graphs illustrate: (a) scale-invariant, and (b) scale-dependent character of box mass distributions $P(m)$. \label{fig7}}
\end{figure}

We examine fractal properties of networks with community structure, the construction of which has been described in Sec.~\ref{SecModel}.  
We found out that the considered network model shows clear fractal properties only near the percolation threshold (see Fig.~\ref{fig6}). The initially power-law-like plot of $N_B$ versus $l_B$, Eq.~(\ref{defdB}) becomes more and more exponential as one moves away from the critical line in the phase diagram (see Fig.~\ref{fig4}). Interestingly, when the networks reveal fractality, all three distributions characterizing: group sizes, $P(s)$, community sizes, $P(c)$, and box masses, $P(m)$, coincide with each other (see Fig.~\ref{fig7}~(a)). Moreover, just like in other fractal complex networks, the box mass distribution does not depend on $l_B$ (see inset in Fig.~\ref{fig7}~(a)). 

On the other hand, when the model loses its fractal properties, although the distributions $P(s)$ and $P(c)$ still overlap, the box mass distributions $P(m)$ distort (see Fig.~\ref{fig7}~(b)). This is due to the fact that the increasing number of connections between modules, which act as shortcuts, destabilizes Song's algorithm. As a result, one giant box is formed, which contains a large number of separate groups/communities. 

Figures~\ref{fig6} and~\ref{fig7}, show the continuous change in model properties as one moves away from the percolation threshold. This observation suggests that a kind of fractal core, slowly getting superstructured by the large amount of the type $B$ edges, may be also present (hidden) in the non-fractal networks far from the threshold. In what follows we show that this is indeed the case.

\subsection{Uncovering the fractal core from the community structure}

\begin{figure}[t]
	\includegraphics[width=0.88\columnwidth]{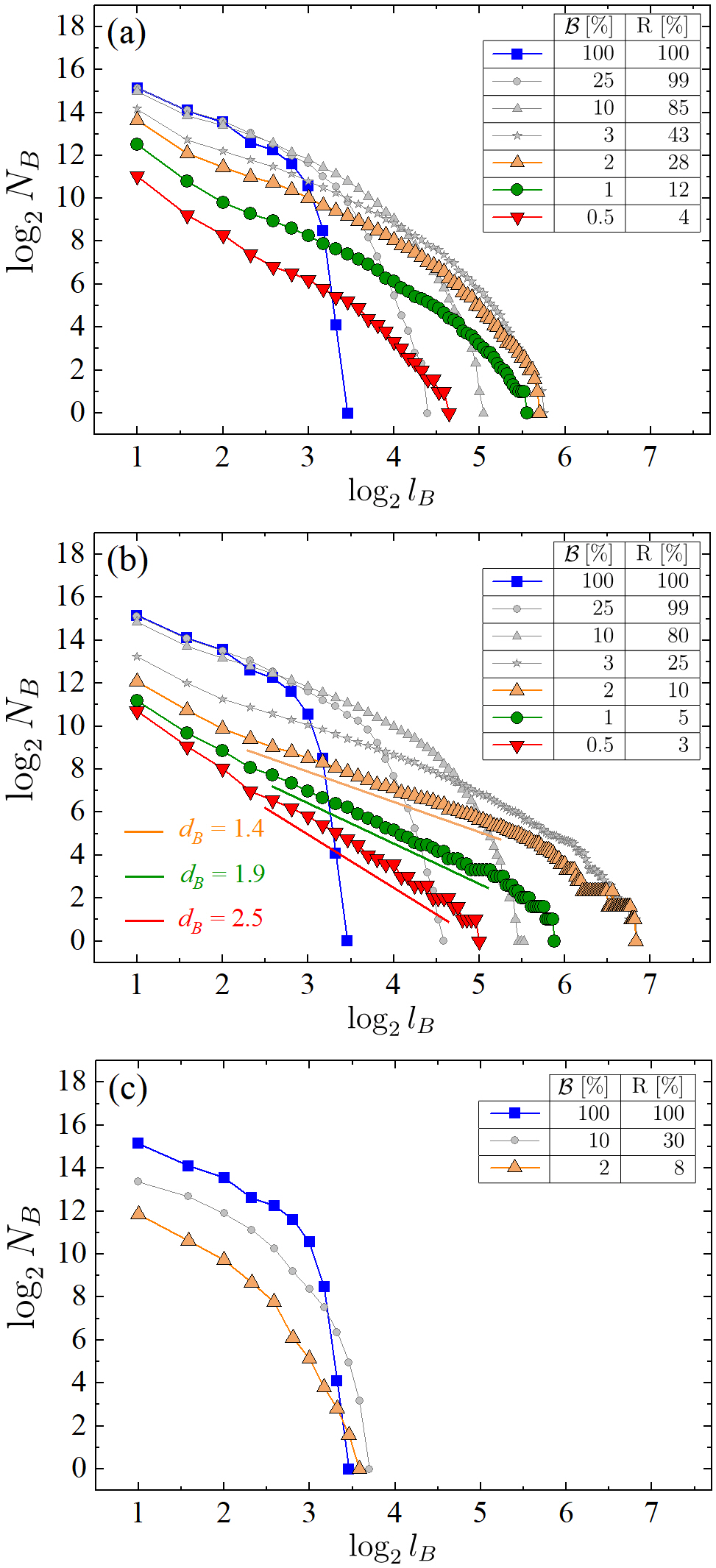}
	\caption{Box counting analysis of the network cores arising from the non-fractal network with $p=0.35$, $a=4$, and $b=1$; cf. Fig.~\ref{fig6}. The following graphs correspond to different edge removal strategies: (a) random, and (b,c) according to decreasing and increasing edge betweenness centrality, respectively. The data series provided correspond to different values of: $\mathcal{B}$ - the relative number of inter-module edges left in the network, and $R$ the relative size of the core, both parameters are given in percentages. \label{fig8}}
\end{figure}

To verify the hypothesis of fractal cores underlying the structure of communities in complex networks, we developed three scenarios for thinning the network by removing type $\mathcal{B}$-edges (i.e., edges between communities identified by the Leiden algorithm that are known to destabilize Song's algorithm and destroy fractality). More specifically, we tested the random removal of edges, as well as the removal of edges according to their relevance, starting with the most and least significant edges, respectively, with relevance measured by the edge betweenness centality (BC) \cite{2001GohPRL, 2006GohPRL} (see Fig.~\ref{fig8}).

Each of the edge removal strategies analyzed leads to splitting the network into smaller and smaller clusters. We focused on the largest connected components of the network diluted in this way. We found that removing $\mathcal{B}$-edges randomly leads to much slower reduction in the relative size $R$ of the largest connected component than their removing according to decreasing or increasing BC. Furthermore, only the scenario: \textit{the highest BC first} leads to the discovery of the macroscopic cores having fractal properties. In the two other scenarios the cores remain non-fractal until the network breaks down into many microscopic components. 

In Fig. ~\ref{fig8} (b), we also see that removing more type $\mathcal{B}$-edges leads to improved exposure of the fractal core over a wider range of box sizes. Interestingly, the box dimension $d_B$ of the fractal core approaches the box dimension of the fractal network that we observe at the percolation threshold Fig.~\ref{fig6} as it becomes more and more exposed. This may mean that our model has some general predefined fractal properties that we are slowly recovering during the iterative process of edge removal.

\section{Summary and concluding remarks}\label{SecSummary}

To summarise, in this paper we propose a hypothesis suggesting that the community structures observed in various complex networks might arise from their concealed fractal characteristics. To assess this hypothesis, we introduce a novel model for evolving networks with community structures, that demonstrates dual scale-invariance, both at the level of node degrees and community sizes. Our findings indicate that, at least within this model, we cannot dismiss the proposed hypothesis. Rationale for this lies in the identification of a fractal core within these networks. Proposed edge removal scheme that reveals those fractal cores refers to the idea of repulsion between hubs \cite{2006SongNatPhys}. This does not mean, however, that the used scheme is the only correct one. Rather, we believe that the method for recovering fractal cores may be network-specific. To justify this belief the concept of BC-maximizing tree-like skeletons of fractal networks can be invoked \cite{2006GohPRL,2007KimPRE}, which were shown to characterize inherently fractal real networks, most of which have a well-defined community structure. This remark indicates the need to perform comprehensive research on real networks with community structure, which may lead to the discovery of hitherto unknown (fractality-driven) universality classes for such networks. Another interesting research direction as a continuation of \cite{Wang_2012,Liu_2019} would be to potentially see if multifractal cores can be detected in complex networks using similar methods.


\section{Acknowledgements}

Research was funded by POB Cybersecurity and Data Science (MS, KM, AF) and POSTDOC PW programmes (PF, MŁ) of Warsaw University of Technology within the Excellence Initiative: Research University (IDUB).

\appendix*
\section{}\label{appendix1}

Below we present a sketch of analytical derivations for the $P(k)$ and $P(s)$ distributions of the network model introduced in Sec.~\ref{SecModel}.

We start with emphasizing that, in the considered model, the time $t$ is measured with respect to the number of nodes $N$ added to the network, i.e. 
\begin{equation}\label{eq0}
	t=\frac{N}{g}-1.
\end{equation}
Relying on the approximation that treats time and node degrees as continuous variables, the time dependence of the node degree $k_i$ can be calculated using the following rate equation: 
\begin{equation}\label{eq1}
	\frac{dk_i}{dt} = \frac{dk_i^A}{dt} + \frac{dk_i^B}{dt},
\end{equation}
where, in a single time step $dt$, the average increase in the node degree resulting from new intra-module edges is:
\begin{equation}\label{eq2}
	\frac{dk_i^A}{dt} = pga\frac{Q_i}{Q}\frac{k_i}{Q_i} = pga\frac{k_i}{Q},
\end{equation}
and the corresponding increase originating from inter-module connections reads:
\begin{equation}\label{eq3}
	\frac{dk_i^{(b)}}{dt} = gb\frac{k_i}{Q},
\end{equation}
where $Q_i$ is the total degree of the group to which the node $i$ belongs and $Q$ is the expected total degree of the whole network, i.e.
\begin{eqnarray}\nonumber
	Q(t)&=&2t\left[pg(a+b)+(1-p)\left({g\choose 2}+gb\right)\right]\\\label{eq4}
	&\stackrel{g=a\!+\!1}{=}& t(a+1)(pa+a+2b).
\end{eqnarray} 

After substituting Eqs.~(\ref{eq2})-(\ref{eq4}) into Eq.~(\ref{eq1}), the rate equation for the node degree reads: 
\begin{equation}\label{eq5}
	\frac{dk_i}{dt} =x\frac{k_i}{t},
\end{equation}
where
\begin{equation}\label{eq6}
	x=\frac{pa+b}{pa+a+2b}.
\end{equation}
This equation can be readily integrated with the initial condition $k_i(t_i)=a+b$, yielding 
\begin{equation}\label{eq6}
	k_i(t) = (a+b)\left(\frac{t}{t_i}\right)^x.
\end{equation}
Finally, by noticing that the time $t_i$ at which the node $i$ enters the network is uniformly distributed in $[0,t]$, the node degree distribution can be obtained from the integral: 
\begin{equation}\label{eq7}
	P(k,t) = \frac{1}{N}\int_0^t\delta[k-k_i(t)]dt_i,
\end{equation}
where $N$, Eq.~(\ref{eq0}), stands for the network size and $\delta$ is the Dirac delta function. Solving the above equation and considering the infinite time limit $t,N\rightarrow\infty$ yields:
\begin{equation}\label{Pk}
	P(k)\simeq k^{-\gamma},
\end{equation} 
with the characteristic exponent given by the model parameters:
\begin{equation}\label{gamma}
	\gamma=1+\frac{1}{x}=2+\frac{a+b}{pa+b}.
\end{equation}

In a similar way, using the continuous-time mean field method (i.e. solving the relevant rate equations for the size and total degree of a group, and then mapping the uniform distribution of group birth times into the group size distribution, cf. Eq.~(\ref{eq2})-(\ref{eq7})), one can show that, in the considered networks, the theoretical group size distribution is also scale-free:
\begin{equation}\label{Ps}
	P(s)\simeq s^{-\eta},
\end{equation} 
with
\begin{equation}\label{eta}
	\eta=1+\frac{pa+a+2b}{p(2a+b)+b}.
\end{equation}

\bibliography{samsel2Refs}

\end{document}